\documentclass[twocolumn, aps, showpacs,preprintnumbers,amsmath,amssymb]{revtex4}

\usepackage{graphicx}
\usepackage{dcolumn}
\usepackage{bm}
\usepackage[colorlinks,citecolor=blue,linkcolor=red]{hyperref}
\usepackage{color}
\usepackage{times}

\begin{document}

\title{Optimal tunneling enhances quantum photovoltaic effect in double quantum dots}

\author{Chen Wang$^{1,}$}\email{wangchen@smart.mit.edu}
\author{Jie Ren$^{2,}$}\email{renjie@lanl.gov}
\author{Jianshu Cao$^{3,1,}$}\email{jianshu@mit.edu}

\address{
$^{1}$Singapore-MIT Alliance for Research and Technology, 1 CREATE Way, Singapore 138602, Singapore\\
$^{2}$Theoretical Division, Los Alamos National Laboratory, Los Alamos, New Mexico 87545, USA \\
$^{3}$Department of Chemistry, Massachusetts Institute of Technology, 77 Massachusetts Avenue, Cambridge, MA 02139, USA
}

\date{\today}

\begin{abstract}
We investigate quantum photovoltaic effect in double quantum dots by applying nonequilibrium quantum master equation. The drastic suppression of the photovoltaic current is observed near the open circuit voltage, which leads to the large filling factor. We find that there always exists an optimal inter-dot tunneling that significantly enhances the photovoltaic current. The maximal output power will also be obtained around the optimal inter-dot tunneling.
Moreover, the open circuit voltage approximately behaves as the product of the eigen-level gap and the Carnot efficiency.
These results suggest a great potential for double quantum dots as efficient photovoltaic devices.
\end{abstract}


\maketitle

\section{Introduction}
As fossil-fuels, the current main energy supplies in our modern society, get scarcer and more expensive, renewable energies become increasingly important and desirable.
To meet this demand, the solar energy, a significant green energy source, attracts  a broad spectrum of attention
from both industrial applications and fundamental researches~\cite{gntiwari1}.
In particular, the photovoltaic effect, firstly discovered by E. Becquerel in $1839$, is a potential promising technology for light harvesting, which converts the inexhaustible sunlight to electricity for performing useful work.

Great efforts have been made to design efficient semiconductor-based solar cells~\cite{magreen1}.
However, the obtained efficiency is still too low to meet human daily needs.
The main reason comes from that the excess excitation energy of the electron-hole pair above the energy gap
will be wasted through thermal phonon emission.
By adding multiple impurity levels, M. Wolf expected the photovoltaic enhancement for the low energy spectrum collection~\cite{mwolf1}.
While Shockley and Queisser suggested that the included impurity would also strengthen the recombination process correspondingly~\cite{wshockley1}, resulting in no improvement of the photovoltaic current.
Moreover, various other proposals have been raised to enhance the solar conversion efficiency~\cite{rtross1,nsano1,skolodinski1,jhwerner1,rbrendel1}.
Recently, quantum dot (QD) emerges as an alternative candidate to fabricate solar cells, due to the ability of enhancing
the photon harvesting via the multi-level structure~\cite{schanyawadee1,kasablon1}.
The novel feature of QD is that by adjusting
the dot size, the energy scale of the excitation gap can be tuned across a wide regime, which extends the absorption spectrum down to the infrared range~\cite{ajnozik1} and makes QD competitive in designing multi-junction solar cells.

Particularly, the influence of the quantum coherence on improving photovoltaic efficiency has been addressed by M. O. Scully et. al~\cite{moscully1,moscully2}.
They studied the photovoltaic cells as quantum heat engine modeled by electronic level systems resonantly coupled to multi-reservoirs with biased temperatures,
which convert incoherent photons to electricity.
Based on the full quantum master equation, which includes the quantum coherence represented as the off-diagonal density matrix elements, the photovoltaic current shows astonishing enhancement compared to the counterpart from population dynamics in the classical limit.
This concept has also been extended to photosynthetic heat engine, which converts solar energy into chemical energy~\cite{kedorfman1,pnalbach1,jscao2,jwu1,jwu2,jwu3}.
From the theoretical view, these generalized engines share the same underlying mechanism.

Considering the importance of quantum coherence in energy conversion for quantum photovoltaic systems, we apply quantum master equation to study the quantum photovoltaic effect in a double quantum dot (DQD) system, which can be also regarded as a donor-acceptor system. In particular, by parallel sandwiching many DQDs between electronic leads, this kind of  nanoscale photovoltaic device could benefit from its flexible
scalability and tunability.
We specially pay attention to the three crucial ingredients of the photovoltaic applications: short circuit current, open circuit voltage and extractable output power, and analyze the ability of the dots to converting photons into electricity. Our results show that there exists an optimal inter-dot tunneling that significantly enhances the quantum photovoltaic current and output power. Moreover, the open circuit voltage approximately behaves as the product of the eigen-level gap and the Carnot efficiency. As a result, the maximal output power will be obtained around the optimal inter-dot tunneling.
The work is organized as follows:
In Sec II., we describe the model of double quantum dots and obtain the solution of the quantum master equation.
In Sec III., we present results and corresponding discussions regarding the quantum photovoltaic effect and current enhancement at optimal tunneling.
A concise summary is given in the final section.

\section{Model and method}

In this section, the model of DQD coupled both
to electron reservoirs and solar environment is first introduced in part A.
Then the quantum master equation is derived in part B,
by assuming the system-reservoir couplings are much weaker than the energy gap of DQD.
Finally in part C, the analytical expressions of steady state electron and photon currents are exhibited.

\subsection{Hamiltonian}

\begin{figure}[tbp]
\includegraphics[scale=0.35]{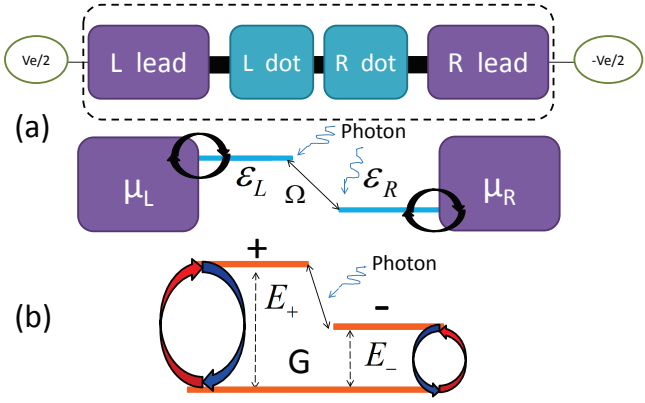}
\vspace{-1.8cm}
\caption{(Color online) (a) Schematic illustration of the DQD device and its photovoltaic dynamics in the real space: the electron hops between the left (right) dot and the left (right) lead, and between the left and right dots; the photon field interacts with the electron population difference of two dots. (b) Scheme of the DQD dynamics in the eigen-space: the photon absorption (emission) assists the excitation (relaxation) between the eigen-state $|-{\rangle}$ and $|+{\rangle}$; the excitation (relaxation) between the ground state $|G{\rangle}$ and the superposition state $|+{\rangle}$ or $|-{\rangle}$ are accompanied by the electron hopping from (to) two electronic leads to (from) the DQD.
}~\label{fig:fig1}
\end{figure}

The photovoltaic system is described by a DQD
coupled to two separate electronic reservoirs [see Fig.~\ref{fig:fig1}(a)], with the total Hamiltonian: $\hat H=\hat H_D+\sum_{v=L,R}(\hat V_v+\hat H_v)+ \hat V_{D-ph}+\hat H_{ph}$. $\hat H_D$ denotes the central DQD by
\begin{eqnarray}~\label{dqd1}
\hat H_D=\epsilon_L\hat d^{\dag}_L\hat d_L+\epsilon_R\hat d^{\dag}_R\hat d_R+\Omega(\hat d^{\dag}_L\hat d_R+\hat d^{\dag}_R\hat d_L),
\end{eqnarray}
where $\hat d^{\dag}_{L(R)}$ creates one electron on the $L(R)$ QD with
energy $\epsilon_{L(R)}$, and $\Omega$ denotes the inter-dot tunneling between $L$ and $R$, which both can be flexibly tuned via gate voltages applied on the dots~\cite{wgvan1}.
Without loss of generality, we consider the strong Coulomb repulsion limit so that the system has three states: the left dot occupied state $|L{\rangle}$, the corresponding right one $|R{\rangle}$, and the ground state $|G{\rangle}$ with both dots empty.
$\hat H_{L(R)}$ depicts the $L(R)$ electronic lead through
$\hat H_v=\sum_k\epsilon_{k,v}\hat c^{\dag}_{k,v}\hat c_{k,v}$,
with $\hat c^{\dag}_{k,v}$ creating one electron with
energy $\epsilon_{k,v}$ and momentum $k$ in the lead $v$.
\begin{eqnarray}~\label{vv1}
\hat V_{v}=\sum_kt_{k,v}\hat d^{\dag}_v\hat c_{k,v}+H.c.
\end{eqnarray}
gives the coupling between the dot $v$ and the lead $v$, which conserves the total electron number and $t_{kv}$ is the system-lead tunneling strength.
When the sun sheds light on the system, the DQD interacts with the photons,
described by
\begin{eqnarray}~\label{VD-ph1}
\hat V_{D-ph}=\sum_qg_q(\hat a_q+\hat a^{\dag}_q)(\hat d^{\dag}_L\hat d_L-\hat d^{\dag}_R\hat d_R),
\end{eqnarray}
where $\hat a^{\dag}_q$ generates one photon with frequency $\omega_q$ in the solar environment
modeled as $\hat H_{ph}=\sum_q\omega_q\hat a^{\dag}_q\hat a_q$,
and $g_q$ is the coupling strength.
Here, we consider that the coupling between the photon environment and the polarization of electron populations on the DQD (instead of the hopping between dots) is the dominant mechanism. This type of electron-photon coupling has been found in DQDs~\cite{mrdelbecq1,mrdelbecq2}, and was already extensively studied for the similar electron-phonon coupling in such systems~\cite{fujisawa1,tbrandes1,gkieblich1,erozbicki1,cwang1a}.
Distinct from the other type of electron-photon coupling 
$\sum_q g_q(\hat{a}_q+\hat{a}^{\dag}_q)(\hat{d}^{\dag}_L\hat{d}_R+\hat{d}^{\dag}_R\hat{d}_L)$ that explicitly describes the photon-assisted tunneling, it seems not obvious that Eq. (3) is able to produce the photovoltaic effect in the local basis. However, as we will show soon, by transforming the system into eigen-space [see also Fig. 1(b)], it is clear that the photon-assisted tunneling emerges with the help of inter-dot tunneling $\Omega$ in Eq. (1). This inter-dot tunnling, on the one hand assists the photovoltaic current, on the other hand diminishes the photovoltaic current. Thus, an optimal inter-dot tunneling will be obtained to enhance the photovoltaic effect.

To investigate the quantum evolution of the system density matrix, it is more convenient to work in the eigen-space of the
DQD by diagonalizing Eq.~(\ref{dqd1}):
\begin{eqnarray}~\label{eigenstate1}
|+\rangle&=&\cos\frac{\theta}{2}|L\rangle+\sin\frac{\theta}{2}|R\rangle,\nonumber\\
|-\rangle&=&-\sin\frac{\theta}{2}|L\rangle+\cos\frac{\theta}{2}|R\rangle,
\end{eqnarray}
which are superpositions of the left and right occupied states,
with $\tan\theta=2\Omega/\Delta$ and $\Delta=\epsilon_L-\epsilon_R$ the inter-dot energy gap.
The corresponding eigen-levels are
\begin{eqnarray}~\label{eigenstate2}
E_{\pm}=\frac{\epsilon_L+\epsilon_R}{2}{\pm}\frac{\sqrt{\Delta^2+4\Omega^2}}{2}.
\end{eqnarray}
The ground state $|G\rangle$ keeps intact.

\subsection{Quantum master equation}

When the interactions of the DQD with the leads and the photon environment  are
weak~\cite{moscully1,moscully2,kedorfman1}, system-reservoir coupling terms in Eq.~(\ref{vv1}) and Eq.~(\ref{VD-ph1}) can be safely treated perturbatively to the second order.
Further under the Born-Markov approximation, the quantum master equation is given by
\begin{eqnarray}~\label{qme1}
\frac{\partial}{{\partial}t}\hat \rho=-i[\hat H_D,\hat \rho]+\mathcal{L}_e[\hat \rho]+\mathcal{L}_p[\hat \rho].
\end{eqnarray}
where $\hat \rho$ denotes the reduced density matrix for the central DQD.
The first term on the right side shows the unitary evolution of the DQD without the actions from two electronic leads and photons.
The second term exhibits decoherence from the dot-lead coupling, given by [see Appendix \ref{appendixA}]
\begin{eqnarray}~\label{le1}
\mathcal{L}_e[\hat \rho]=&\sum_{v;a=\pm}&\frac{\gamma^a_vd_v^{a}}{2\hbar}\bigg\{
\big(1-f_v(E_a)\big)\left[|G{\rangle}{\langle}a|\hat \rho,\hat d^{\dag}_v\right]\nonumber\\
&&+f_v(E_a)\left[|a{\rangle}{\langle}G|\hat \rho,\hat d_v\right]\bigg\}+H.c. \;.
\end{eqnarray}
$\gamma^a_v=2\pi\sum_{k}|t_{k,v}|^2\delta(\epsilon_{k,v}-E_a)$ denotes the coupling energy between the superposition state $|a\rangle$ ($|+\rangle$ or $|-\rangle$) and the lead $v$. In the following, we assume $\gamma^+_v=\gamma^-_v=\gamma_v$ and set $\gamma_v$ as constant in the wide band limit.
The hopping matrix element $d_{v}^a={\langle}G|\hat d_v|a{\rangle}$, originating from $\hat d_v(-\tau)=\sum_{\omega=E_{\pm}}e^{i{\omega}\tau/\hbar}d_{v}^a|G{\rangle}{\langle}a|+H.c.$,  describes
the electron transfer from the superposition state on DQD to the lead $v$.
$f_v(E_a)=1/(\exp[\beta_v(E_a-\mu_v)]+1)$ is the Fermi-Dirac distribution in the $v$ lead with $\mu_v$ the corresponding chemical potential and $\beta_v=1/(k_BT_v)$ the inverse temperature.
It should be clarified that the expression of Eq.~(\ref{le1}) is based on $E_->0$, which is equivalent to $\epsilon_L\epsilon_R>\Omega^2$.
On the contrary $E_{-}<0$ ($\epsilon_L\epsilon_R<\Omega^2$), it only needs exchange $f_v(E_-)$ with $1-f_v(E_-)$ in Eq.~(\ref{le1}).
When including the external voltage bias,  we conventionally set $\mu_{L(R)}=\mu_0\pm eV_e/2$ with  $\mu_{0}=(\epsilon_L+\epsilon_R)/{2}$.
This enables us to study the current-voltage characteristic of the double quantum dots, which is a crucial ingredient
to design the photovoltaic devices~\cite{gmmasters1}.

The third term depicts the effect of the photon environment on the DQD, shown as [see Appendix \ref{appendixA}]
\begin{eqnarray}~\label{lp1}
\mathcal{L}_p[\hat \rho]&=&\frac{\gamma_pQ_{+-}}{2\hbar}\left\{\big(1+n(\Lambda)\big)[\hat \sigma_-\hat \rho,\hat Q]
+n(\Lambda)[\hat \sigma_+\hat \rho,\hat Q]\right\}\nonumber\\
&&+H.c. \;,
\end{eqnarray}
where $Q_{+-}=\langle+|\hat Q|-\rangle$, $\hat \sigma_{\pm}=|{\pm}{\rangle}{\langle}{\mp}|$ and $\hat Q=\hat d^{\dag}_L\hat d_L-\hat d^{\dag}_R\hat d_R$ describes the population polarization on the DQD.
$\Lambda=E_+-E_-=\sqrt{\Delta^2+4\Omega^2}$ denotes the energy gap of two eigen-levels, $\gamma_p=2\pi\sum_k|g_k|^2\delta(\omega_k-\omega)$ is the coupling energy strength of the photon environment,
and $n(\Lambda)=1/[\exp{(\beta_p\Lambda)}-1]$ is the Bose-Einstein distribution of the photon environment with $\beta_p$ the inverse temperature of the sun. Clearly, only the photons with energy resonant with the eigen-level gap $\Lambda$ will be absorbed.

Eqs.~(\ref{le1}) and~(\ref{lp1}) show that eigen-states $|\pm\rangle$ of DQD are mainly responsible
for the quantum transport, which is also similarly illustrated in Ref.~\cite{crxu1}.
To expose explicitly the physical picture of the photon-assisted transport, we re-express the electron-photon coupling Eq.~(\ref{VD-ph1}) in eigen-state basis
as $\hat V_{D-ph}=\sum_qg_q(\hat a_q+\hat a^{\dag}_q)(\cos\theta\hat \tau_{z}-\sin\theta\hat \tau_{x})$, with $\hat \tau_z=|+\rangle\langle+|-|-\rangle\langle-|$
and $\hat \tau_x=\hat \tau_++\hat \tau_-=|+\rangle\langle-|+|-\rangle\langle+|$.
The first term on the right side of $\hat V_{D-ph}$ is trivial, since it is commutative with the $\hat H_D$.
While for the second term, it appears as $-\sum_q\sin\theta{g_q}(\hat a^{\dag}_q\hat \tau_-+\hat \tau_+\hat a_q)$ under the rotating-wave approximation.
This clearly suggests that the electron hopping between $|\pm\rangle$ is assisted by the photon absorption and emission [see Fig.~\ref{fig:fig1}(b)],
which makes indispensable contribution to the appearance of quantum photovoltaic effect in the DQD system.
Moreover, it should be noted that the evolution equation of the DQD density matrix at Eq.~(\ref{qme1}) has no classical correspondence.
This means no electron or photon current will exhibit by studying the corresponding
population dynamics under local basis.

\subsection{Electron and photon current}

In the Liouville space, the density matrix of the double quantum dots is expressed in the vector form
$|\mathbb{P}\rangle=(\rho_{GG},\rho_{LL},\rho_{RR},\rho_{LR},\rho_{RL})^{T}$, with $\rho_{ij}={\langle}i|\hat \rho|j{\rangle}$.
Then the evolution equation is re-expressed as [see Appendix \ref{appendixA}]: 
\begin{eqnarray}~\label{leq1}
\frac{\partial}{{\partial}t}|\mathbb{P}\rangle=\mathbb{L}|\mathbb{P}\rangle,
\end{eqnarray}
where $\mathbb{L}$ is the matrix form of Liouville superoperator.
The steady state solution is obtained through $\mathbb{L}|\mathbb{P}^{ss}\rangle=0$, with $|\mathbb{P}^{ss}\rangle$ the steady state density vector.
Define the direction from right to left as positive, the photovoltaic current is obtained [see Appendix \ref{appendixB}], as
\begin{eqnarray}~\label{Ie1}
I_e/e=\Gamma_{L}\rho^{ss}_{LL}-\Gamma_{GL}\rho^{ss}_{GG}+2\Theta_{GL}\textrm{Re}[\rho^{ss}_{LR}],
\end{eqnarray}
where
$\Gamma_L=\frac{\gamma_L}{\hbar}\left(\cos^2\frac{\theta}{2}[1-f_L(E_+)]+\sin^2\frac{\theta}{2}[1-f_L(E_-)]\right)$ denotes the electron hopping rate from the left dot to the left lead; $\Gamma_{GL}=\frac{\gamma_L}{\hbar}\left(\cos^2\frac{\theta}{2}f_L(E_+)+\sin^2\frac{\theta}{2}f_L(E_-)\right)$ is the reverse-process rate from the left lead to the left dot; $\Theta_{GL}=\frac{\gamma_L\sin\theta}{4\hbar}\left(f_L(E_-)-f_L(E_+)\right)$ depicts the relaxation rate from the quantum coherent state between the left and right dots to the ground state by emitting an electron into the left lead. This process is a pure quantum effect and gives the positive contribution to the right-to-left current.
Similarly, the photon current absorbed from the solar environment can be also obtained as [see Appendix \ref{appendixB}]
\begin{equation}\label{ip1}
I_p=-\frac{\gamma_p\sin\theta}{2\hbar}\left(\sin\theta(\rho^{ss}_{LL}+\rho^{ss}_{RR})+2[1+2n(\Lambda)]\textrm{Re}[\rho^{ss}_{LR}]\right).
\end{equation}

Eq.~(\ref{Ie1}) and Eq.~(\ref{ip1}) imply that quantum coherence, manifested by $\rho^{ss}_{LR}$, is crucial to correctly describe the current.
Moreover, the factor of $\sin\theta$ in $I_p$ shows that photon current vanishes at $\theta=0$, i.e. at $\Omega=0$.
Accordingly, in the absence of inter-dot electron tunneling, $|L(R){\rangle}$ state keeps equilibrium with its own reservoir
under the relation $\rho^{ss}_{LL(RR)}/\rho^{ss}_{GG}=\exp{(-\beta_{L(R)}\Delta/2)}$, which readily leads to $I_e=0$ since the last contribution from the quantum coherence vanishes  when $\Omega=0$.
On the opposite limit when the inter-dot coupling $\Omega$ becomes large, the electron population polarization of the DQD will be small so that the electron-photon coupling becomes rather weak [see Eq.~(\ref{VD-ph1})]. Moreover, increasing $\Omega$ will enhance the back-tunneling current from left to right. As a result, the photovoltaic current will be severely suppressed at large $\Omega$.
Thus, it is naturally to expect the maximal photovoltaic behavior in the intermediate tunneling regime.

\section{Result and Discussion}

\subsection{I-V curves}

\begin{figure}[tbp]
\includegraphics[scale=0.45]{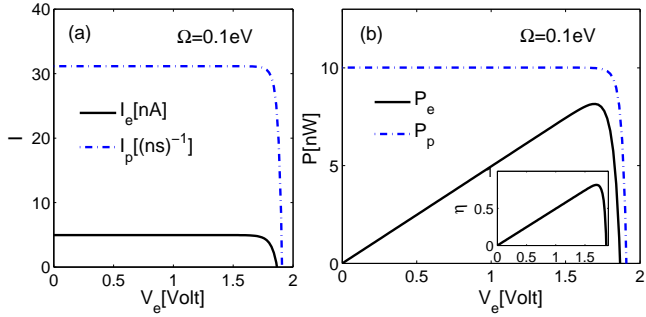}
\vspace{-5mm}
\caption{(Color online) Currents (photovoltaic current $I_e$ and photon flow $I_p$)
and energy power (photovoltaic power $P_s$ and solar power $P_p$) as functions of external voltage bias $V_e$.
Other parameters are $\epsilon_L=3$~eV, $\epsilon_R=1$~eV,
$\gamma_L=\gamma_R=\gamma_p=0.1$~eV, $T_L=T_R=300$~K, $T_p=6000$~K.}~\label{fig:fig2}
\end{figure}

The current-voltage characteristic (I-V curve) is crucial for analyzing the quantum photovoltaic effect,
in which the short circuit current, open circuit voltage and photovoltaic power can be explicitly identified~\cite{hychen1,mshalom1,pvkamat1,jtang1}.
We first investigate the photovoltaic current and the output power
in Fig.~\ref{fig:fig2}.
The temperatures of both the left and right leads are set to the room temperature.
For solar photons, the temperature is chosen by $T_p=6000$~K as traditionally described~\cite{meinax1}. 
As shown in Fig.~\ref{fig:fig2}(a), when the voltage bias is turned on but small, the electron current keeps nearly the same strength as the short circuit current $I^{sc}_e$.
However, when the voltage approaches the open circuit voltage $V_{oc}$, the electron current is sharply  suppressed down to zero.
Hence, the DQD has a high filling factor, which is crucial for high efficiency~\cite{ltdou1,zche1,zche2}.
The similar feature has been described in other photovoltaic realizations~\cite{moscully2,meinax1,kedorfman1}, considered as a key element to design efficient photovoltaic devices.
In the recent studies regarding the cavity quantum electrodynamics system~\cite{crxu1} and organic Heterojunction~\cite{meinax1}, the photovoltaic current is exhibited as $I_e{\sim}1$~pA and $I_e{\sim}10$~pA, respectively.
It is much smaller than the present case with nA current scale. This implies that the DQD is an promising candidate serving as the basis of the photovoltaic application.

The behavior of the photon current with the variation of voltage is similar to the electron current,
which also exhibits large suppression near the terminal voltage.
However, the terminal voltage is larger than that ($V_{oc}$) for the electron current [see Fig.~\ref{fig:fig2}(a)]. This is understandable as follows: With finite $\Omega$, the dot system has electron current from left to right under positive voltage
in absence of the electron-photon interaction.
After the electron-photon coupling is included, the photon absorption by quantum dots generates the electron current against the voltage bias,
originating from the quantum photovoltaic effect.
Therefore, the electron current is composed by two competing sources:
(i) intrinsic tunneling between QDs generates downhill current under positive voltage, which gives negative contribution to the electron current,
and (ii) photon-generated uphill current gives crucial positive contribution.
Before the vanishing of the photon current, the photon-generated electron current will be completely eliminated by that from intrinsic inter-dot tunneling at $V_{oc}$, which gives the discrepancy between two terminal voltages.

The photovoltaic (output) and solar (input) powers are studied in Fig.~\ref{fig:fig2}(b).
In the small voltage bias regime, the photovoltaic power $P_e=I_e{\cdot}V_e$ is proportional to $V_e$, until reaching a maximal power,
since the electron current $I_e$ keeps almost constant.
As the voltage reaches $V_{oc}$, the power suddenly drops to zero, due to the drastic diminishing of the current at $V_{oc}$.
For the solar power $P_p=I_p{\cdot}\Lambda$, it is steady at the beginning,
and then decays fast near the terminal voltage, which is consistent with the behavior of $I_p$.
The maximum quantum efficiency $\eta=P_e/P_p$ of the DQD engine is then obtained near $V_{oc}$, as plotted inset in Fig.~\ref{fig:fig2}(b).
This behavior is similar to photovoltaic power and the maximal value is nearly $80\%$.


\begin{figure}[tbp]
\vspace{-0.5cm}
\includegraphics[scale=0.50]{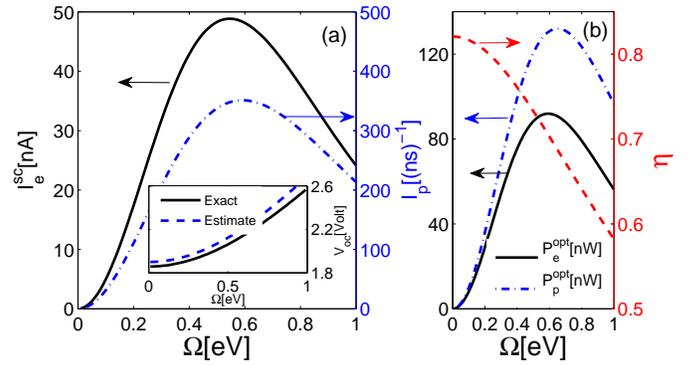}
\vspace{-0.5cm}
\caption{(Color online) (a) Short circuit electron current and photon current at $V_e=0$,
and inset is the comparison of numerically exact $V_{oc}$ from Eq.~(\ref{Ie1}) and the
approximation estimated from Eq.~(\ref{Voc1});
(b) optimal output power $P^{\textrm{opt}}_e=\max\{I_e{\cdot}V\}$, corresponding input power $P^{\textrm{opt}}_p=I_p{\cdot}\Lambda$,
and the efficiency $\eta=P^{\textrm{opt}}_e/P^{\textrm{opt}}_p$, as functions of electron tunneling strength $\Omega$.
Other parameters are the same as those in Fig.~\ref{fig:fig2}.
}~\label{fig:fig3}
\end{figure}

\subsection{Effects of inter-dot quantum tunneling}

Fig.~\ref{fig:fig3}(a) shows the effect of the tunneling on the short circuit current $I^{sc}_e$ at $V_e=0$ in a large scale.
In the weak tunneling regime, the photovoltaic current arises quickly with the increasing tunneling ($I^{sc}_e\sim\Omega^2$), which is also observed in
Fig.~\ref{fig:fig2}.
As the tunneling strength reaches the moderate regime, the electron current peaks at $\Omega\approx0.55$~eV.
After the peak, the current shows monotonic decay.
For the behavior of the photon current, it is similar to the electron current, except for the magnitude difference.
As we discussed above, in absence of inter-dot tunneling, two DQs are decoupled and no photon will be pumped into the dots to generate uphill current,
which is clearly exhibited in Eq.~(\ref{ip1}). Therefore, to obtain photovoltaic effect, finite $\Omega$ is necessary.
In the opposite direction of strong tunneling, the population polarization is very small,
and photons can be hardly  pumped into the system due to the suppressed electron-photon interaction shown in Eq.~(\ref{VD-ph1}).
Moreover, the tunneling also deteriorates the generation of the photovoltaic current.
Hence, it is expected there will exist an optimal tunneling to maximize the photovoltaic current, which is explicitly shown in Fig.~\ref{fig:fig3}(a).

The open circuit voltage with varying tunneling strength is also investigated in the inset of Fig.~\ref{fig:fig3}(a) (solid line), where
$V_{oc}$ shows monotonic  behavior with increasing $\Omega$ that qualitatively coincides with the behavior of the
eigen-level gap $\Lambda=\sqrt{\Delta^2+4\Omega^2}$.
This can be understood as follows:
When the inter-dot tunneling $\Omega$ is weak, it is known that $\sin\theta{\approx}0~(\cos\theta{\approx}1)$ so that $|+{\rangle}$ only effectively connects to the left lead and $|-{\rangle}$ effectively couples with the right lead [see Eqs.~(\ref{coa1},\ref{coa2},\ref{coa3},\ref{coa4})].
Besides, the eigen-levels $|+\rangle$ and $|-\rangle$ are nearly uncoupled since they become orthogonal to each other.
The tunneling between them is mainly assisted by the photon-induced excitation and relaxation.
Hence at the open circuit voltage, considering electron pump from the right ($|-\rangle$) to the left ($|+\rangle$) is balanced by the reverse action,
we have the detailed balance relation:
\begin{equation}
\frac{f_L(E_+)}{1-f_L(E_+)}{\times}\frac{1+n_p(\Lambda)}{n_p(\Lambda)}{\times}\frac{1-f_R(E_-)}{f_R(E_-)}=1,
\end{equation}
where the rate from the left lead to the right one is proportional to $f_L(E_+)[1+n_p(\Lambda)][1-f_R(E_-)]$ while the reverse rate from right to left is proportional to $[1-f_L(E_+)]n_p(\Lambda)f_R(E_-)$. This detailed balance relation finally gives us
\begin{equation}~\label{Voc1}
V_{oc}=\frac{\Lambda}{e}\left(1-\frac{T_{0}}{T_p}\right),
\end{equation}
where $T_{0(p)}$ denotes the electronic reservoirs (solar environment) temperature and $1-{T_{0}}/{T_p}$ is the ideal Carnot efficiency.
This rough estimation qualitatively agrees with the numerical exact result in the inset of Fig.~\ref{fig:fig3}(a),
and the slight deviation comes from the weak inter-dot tunneling, which reduces $V_{oc}$ compared to the ideal one at Eq.~(\ref{Voc1}).
From these results, it is interesting to find that below the optimal tunneling ($\Omega\approx0.55$~eV in our case),
both the photovoltaic current and voltage are enhanced by the tunneling strength.
Thus, the best operation regime is around the optimal tunneling, where the maximum output power will be obtained.
This feature is explicitly shown in Fig.~\ref{fig:fig3}(b).
However, the photovoltaic efficiency corresponding to the maximal extractable output power is not the largest,
which shows monotonic decay.
This provides useful guidance to optimize the quantum photovoltaic effect.

\begin{figure}[tbp]
\vspace{-0.5cm}
\includegraphics[scale=0.42]{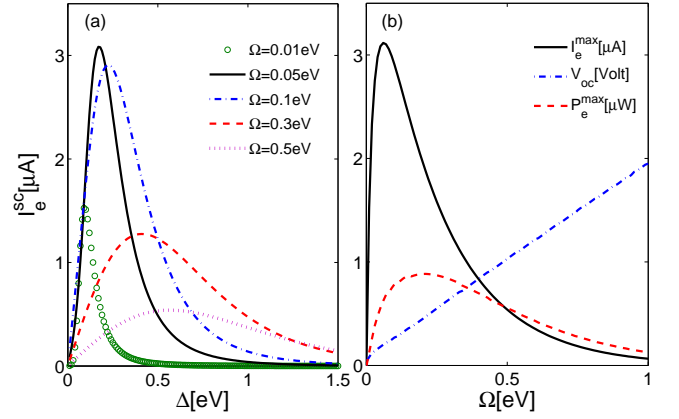}
\vspace{-1.8cm}
\caption{(Color online) (a) Short circuit current as a function of $\Delta$ with different $\Omega$;
(b) Maximum of the photovoltaic current, the corresponding open circuit voltage and output power, under various $\Omega$.
Other parameters are the same as those in Fig.~\ref{fig:fig2}.}~\label{fig:fig4}
\end{figure}

\subsection{Global optimal performance}

Next, we study the effect of the inter-dot energy gap $\Delta$ on the photovoltaic current in Fig.~\ref{fig:fig4}(a).
For arbitrary tunneling strength, there always exists an optimal gap to maximize the current.
Moreover, the overall profiles are similar:
the current firstly arises with increasing $\Delta$, and then it decays monotonically
after reaching the maximum.
However, the differences are also apparent.
For weak tunneling, i.e. $\Omega=0.01$~eV, The value of the peak is small, around $1.5$~${\mu}$A.
As the tunneling is strengthened, this value becomes large, i.e. $I^{\max}_e\approx3$~${\mu}$A.
When the tunneling is further  increased, the current again becomes weak.
Besides, the peak is broadened with increasing $\Omega$.
Based on the results of Fig.~\ref{fig:fig4}(a),
we extract the maximum values of the current ($I^{\max}_e$) and investigate their dependence on the tunneling strength,
shown in Fig.~\ref{fig:fig4}(b). The global summit appears at $\Omega\approx0.08$~eV, which corresponds to the gap of two excited states $\Lambda\approx0.3$~eV.
Hence the central frequency of the absorption photons is in the infrared regime~\cite{kasablon1},
and the maximum value of the current can be as large as $3$ ${\mu}$A.
It shows competitive improvement by comparing with photovoltaic current in other photocell unit, 
i.e. $I_{sc}{\sim}1$~pA in Ref.~\cite{crxu1} and $I_{sc}{\sim}10$~pA in Ref.~\cite{meinax1}.

For the open circuit voltage corresponding to the maximum short circuit current, it changes almost linearly with $\Omega$
(we also find the excellent linear relation of $V_{oc}$ with $\Lambda$)~\cite{wjyoon1,mcscharber1},
which is quite different from that in Fig.~\ref{fig:fig3}(b).
The difference mainly comes from the different flexibility of the energy bias $\Delta$.
For the formal case in Fig.~\ref{fig:fig3}(b), the energy bias is fixed with $\Delta=2$~eV and does not change with the variation of $\Omega$.
While for the present case, the maximum electron current shows the global picture in the parameter space of $\Delta$ and $\Omega$, where $\Delta$ is adjusted with varying $\Omega$.
We also investigate the maximum power, defined as $P^{\max}_e=I^{\max}_e{\cdot}V_{oc}$.
It also shows the peak effect with the optimal tunneling, but the optimal point deviates from that for the photovoltaic current.
As is well-known, over $50\%$ of the solar energy is below the visible light spectrum~\cite{gfmoore1}.
Therefore, our results suggest that it is meaningful to use the DQD as one basis for the design of efficient solar energy harvesters.

\section{Conclusion}

In summary, we have studied the quantum photovoltaic effect in a DQD system weakly coupled to electronic leads and solar environment
by applying the quantum master equation.
Three main ingredients of photovoltaic effect: short circuit current, open circuit voltage and output power, have been analyzed in detail.
As the voltage bias approaches open circuit voltage ($V_{oc}$), the electron current is strongly suppressed to zero,
implying the high fill factor.
In comparison, the photon current is eliminated at a larger terminal voltage.
This discrepancy mainly originates from that the photovoltaic current is composed by
two competing sources, one from the photon-generated uphill electron current against the potential bias,
and the other from the voltage bias driving the electron current along the potential gradient.
When the photovoltaic current disappears, these two sources induced currents are equal, resulting in the finite photon current.
Moreover, the photovoltaic current and power are much larger than other recently studied nano-junction photovoltaic systems,
which is crucial for designing photovoltaic devices.

The influence of the inter-dot tunneling strength on the photovoltaic current is investigated.
The optimal tunneling to maximize the photovoltaic current has been found in the intermediate regime,
of which the character should be intrinsic in this kind system.
Whereas the open circuit voltage increases monotonically with the increasing tunneling,
which can be qualitatively described by $V_{oc}{\sim}\Lambda(1-T_{0}/T_p)$, based on the detailed balance condition.
The global optimal tunneling to achieve the maximal photovoltaic current and power has been also exhibited,
with the central frequency of absorption photons in the infrared regime.
We believe that these results provide theoretical basis for promising photovoltaic applications of  double quantum dots.

\begin{appendix}

\section{Quantum Master Equation under Counting Field}\label{appendixA}

To derive the electron current and the photon flow, we usually include the counting field as in the method of full counting statistics~\cite{ymblanter1,lslevitov1,lslevitov2,mesposito1}.
Here, we count the electron number $\hat{N}_L=\sum_k\hat{c}^{\dag}_{k,L}\hat{c}_{k,L}$ on the left fermion reservoir
and the photon number $\hat{N}_p=\sum_k\hat{a}^{\dag}_k\hat{a}_k$ in the solar environment.
The Hamiltonian of the whole system is modified to~\cite{cwang1a}
\begin{eqnarray}
\hat{H}_{\chi}&=&e^{i(\hat{N}_L\chi_e+\hat{N}_p\chi_p)/2}\hat He^{-i(\hat{N}_L\chi_e+\hat{N}_p\chi_p)/2}\\
&=&\hat{H}_D+\sum_{v=L,R}(\hat{V}^{\chi_e}_{v}+\hat{H}_v)+\hat{V}^{\chi_p}_{D-ph}+\hat{H}_{ph},\nonumber
\end{eqnarray}
where $\chi=(\chi_e,\chi_p)$ count the currents transferring into the corresponding reservoirs, and the system-bath interactions are modified to
\begin{eqnarray}
\hat{V}^{\chi_e}_v&=&\sum_{k,v}t_{k,v}e^{-i\chi_e\delta_{v,L}/2}\hat{d}^{\dag}_v\hat{c}_{k,v}+H.c.,\nonumber\\
\hat{V}^{\chi_p}_{D-ph}&=&\sum_qg_q(\hat{a}_qe^{-i\chi_p/2}+\hat{a}^{\dag}_qe^{i\chi_p/2})(\hat{d}^{\dag}_L\hat{d}_L-\hat{d}^{\dag}_R\hat{d}_R),\nonumber
\end{eqnarray}
with $\delta_{\alpha,\beta}=1$ if $\alpha=\beta$, otherwise $\delta_{\alpha,\beta}=0$.
Following the standard procedure treated in quantum master equation including counting field up
to the second order~\cite{mesposito1,cwang1a,tyuge1a},
the dissipator from the QD-electron reservoir is derived as
\begin{eqnarray}
\mathcal{\hat{L}}_e[\hat{\rho}_{\chi}]&&=\sum_{v,a}\frac{\gamma^a_vd_{v,Ga}}{2\hbar}\{
f_v(E_a)e^{-i\chi_e\delta_{v,L}}(\hat{d}^{\dag}_v\hat{\rho}|G{\rangle}{\langle}a|+H.c.)\nonumber\\
&&+(1-f_v(E_a))e^{i\chi_e\delta_{v,L}}(\hat{d}_v\hat{\rho}|a{\rangle}{\langle}G|+H.c.)\nonumber\\
&&-([(1-f_v(E_a))\hat{d}^{\dag}_v|G{\rangle}{\langle}a|\hat{\rho}+f_v(E_a)\hat{d}_v|a{\rangle}{\langle}G|\hat{\rho}]\nonumber\\
&&+H.c.)\},
\end{eqnarray}
with $v=L,R$ and $a=\pm$.
It will naturally reduce to Eq.~(\ref{le1}) when $\chi_e=0$.
And the Liouville operator from the dot-photon coupling is shown as
\begin{eqnarray}
\mathcal{\hat{L}}_p[\hat{\rho}_{\chi}]&&=\frac{\gamma_pQ_{+-}}{2\hbar}\{
n(\Lambda)e^{-i\chi_p}(\hat{Q}\hat{\rho}|-\rangle\langle+|+H.c.)\nonumber\\
&&+(1+n(\Lambda))e^{i\chi_p}(\hat{Q}\hat{\rho}|+\rangle\langle-|+H.c.)\nonumber\\
&&-([(1+n(\Lambda))\hat{Q}|-\rangle\langle+|\hat{\rho}+n(\Lambda)\hat{Q}|+\rangle\langle-|\hat{\rho}]\nonumber\\
&&+H.c.)\}.
\end{eqnarray}
When $\chi_p=0$, it returns back to Eq.~(\ref{lp1}) consistently.
Then the quantum master equation under counting field is described by
\begin{eqnarray}
\frac{\partial}{{\partial}t}\hat{\rho}_{\chi}=-i[\hat{H}_D,\hat{\rho}_{\chi}]+\mathcal{\hat{L}}_e[\hat{\rho}_{\chi}]+\mathcal{\hat{L}}_p[\hat{\rho}_{\chi}].
\end{eqnarray}
Furthermore, in the Liouville space the reduced density matrix of the DQD system is expressed as vector form
$|\mathbb{P}_{\chi}\rangle=(\rho_{GG},\rho_{LL},\rho_{RR},\rho_{LR},\rho_{RL})^{T}$, with $\rho_{ij}={\langle}i|\hat{\rho}_{\chi}|j{\rangle}$.
Hence, the corresponding evolution equation of the DQD density matrix is given by
\begin{eqnarray}
\frac{\partial}{{\partial}t}|\mathbb{P}_{\chi}\rangle=\mathbb{L}_{\chi}|\mathbb{P}_{\chi}\rangle,
\label{eq:leq1}
\end{eqnarray}
with $\mathbb{L}_{\chi}=\mathbb{L}^e_{\chi_e}+\mathbb{L}^p_{\chi_p}$. When $\chi_e=\chi_p=0$, Eq.~(\ref{eq:leq1}) is just simplified back to Eq.~(\ref{leq1}) with $|\mathbb{P}_{\chi}\rangle$ reducing to $|\mathbb{P}\rangle$ and $\mathbb{L}_{\chi}$ reducing to $\mathbb{L}$.
Here $\mathbb{L}^e_{\chi_e}$ describes the superoperator for the electron leads induced decoherence as
\begin{widetext}
\begin{eqnarray}
\mathbb{L}^e_{\chi_e}=
\begin{pmatrix}
-(\Gamma_{GL}+\Gamma_{GR}) & \Gamma_Le^{i\chi_e} & \Gamma_R & \Theta_{GL}e^{i\chi_e}+\Theta_{GR} & \Theta_{GL}e^{i\chi_e}+\Theta_{GR}\\
\Gamma_{GL}e^{-i\chi_e} & -\Gamma_L & 0 & -\Theta_{GL}+i\Omega & -\Theta_{GL}-i\Omega\\
\Gamma_{GR} & 0 & -\Gamma_R & -\Theta_{GR}-i\Omega & -\Theta_{GR}+i\Omega\\
\Gamma^{e}_{\chi_e} & -\Theta_{GR}+i\Omega & -\Theta_{GL}-i\Omega & -\frac{\Gamma_L+\Gamma_R}{2}-i\Delta & 0\\
\Gamma^{e}_{\chi_e} & -\Theta_{GR}-i\Omega & -\Theta_{GL}+i\Omega & 0 & -\frac{\Gamma_L+\Gamma_R}{2}+i\Delta\\
\end{pmatrix},
\end{eqnarray}
\end{widetext}
where $\Delta=\epsilon_L-\epsilon_R$, and the other renormalized parameters are explicitly given by
\begin{equation}
\Gamma_{GL}=\frac{\gamma_L}{\hbar}(\cos^2\frac{\theta}{2}f_L(E_+)+\sin^2\frac{\theta}{2}f_L(E_-)),
~\label{coa1}
\end{equation}
\begin{equation}
\Gamma_{GR}=\frac{\gamma_R}{\hbar}(\sin^2\frac{\theta}{2}f_R(E_+)+\cos^2\frac{\theta}{2}f_R(E_-)),
~\label{coa2}
\end{equation}
\begin{equation}
\Gamma_{L}=\frac{\gamma_L}{\hbar}(\cos^2\frac{\theta}{2}[1-f_L(E_+)]+\sin^2\frac{\theta}{2}[1-f_L(E_-)]),
~\label{coa3}
\end{equation}
\begin{equation}
\Gamma_{R}=\frac{\gamma_R}{\hbar}(\sin^2\frac{\theta}{2}[1-f_R(E_+)]+\cos^2\frac{\theta}{2}[1-f_R(E_-)]),
~\label{coa4}
\end{equation}
\begin{equation}
\Theta_{GL(GR)}=\frac{\sin\theta\gamma_{L(R)}}{4\hbar}(f_{L(R)}(E_-)-f_{L(R)}(E_+)),
\end{equation}
\begin{eqnarray}
\Gamma^e_{\chi_e}&=&\frac{\sin\theta}{4\hbar}(\gamma_L[f_L(E_+)-f_L(E_-)]e^{-i\chi_e}\nonumber\\
&&+\gamma_R[f_R(E_+)-f_R(E_-)]).
\end{eqnarray}

While $\mathbb{L}^p_{\chi_p}$ accounts for the electron-photon interaction, shown as
\begin{eqnarray}
\mathbb{L}^p_{\chi_p}=
\begin{pmatrix}
0 & 0 & 0 & 0 & 0\\
0 & -\Gamma^p_{\chi_p} & 0 & -\Gamma^p_{L,\chi_p} & -\Gamma^p_{L,\chi_p}\\
0 & 0 & -\Gamma^p_{\chi_p} & -\Gamma^p_{R,\chi_p} & -\Gamma^p_{R,\chi_p}\\
0 & -\Theta^1_{\chi_p} & -\Theta^2_{\chi_p} & -\Theta^3_{\chi_p} & 0\\
0 & -\Theta^1_{\chi_p} & -\Theta^2_{\chi_p} & 0 & -\Theta^3_{\chi_p}\\
\end{pmatrix},
\end{eqnarray}
with the elements
\begin{eqnarray}
\Gamma^p_{\chi_p}&=&\frac{\gamma_p\sin^2\theta}{2\hbar}([1+2n(\Lambda)]-n(\Lambda)e^{-i\chi_p} \nonumber\\
&&-[1+n(\Lambda)]e^{i\chi_p}),\\
\Gamma^p_{L,\chi_p}&=&\frac{\sin\theta\gamma_p}{2\hbar}[\sin^2\frac{\theta}{2}[1+n(\Lambda)](1-e^{i\chi_p})\nonumber\\
&&-\cos^2\frac{\theta}{2}n(\Lambda)(1-e^{-i\chi_p})], \\
\Gamma^p_{R,\chi_p}&=&\frac{\sin\theta\gamma_p}{2\hbar}(\cos^2\frac{\theta}{2}[1+n(\Lambda)](1-e^{i\chi_p})\nonumber\\
&&-\sin^2\frac{\theta}{2}n(\Lambda)(1-e^{-i\chi_p})), \\
\Theta^1_{\chi_p}&=&\frac{\sin\theta\gamma_p}{2\hbar}(\cos^2\frac{\theta}{2}[1+n(\Lambda)](1+e^{i\chi_p})\nonumber\\
&&-\sin^2\frac{\theta}{2}n(\Lambda)(1+e^{-i\chi_p})), \\
\Theta^2_{\chi_p}&=&\frac{\sin\theta\gamma_p}{2\hbar}(\sin^2\frac{\theta}{2}[1+n(\Lambda)](1+e^{i\chi_p})\nonumber\\
&&-\cos^2\frac{\theta}{2}n(\Lambda)(1+e^{-i\chi_p})), \\
\Theta^3_{\chi_p}&=&\frac{\sin^2\theta\gamma_p}{2\hbar}([1+2n(\Lambda)]+n(\Lambda)e^{-i\chi_p}\nonumber\\
&&+[1+n(\Lambda)]e^{i\chi_p}).
\end{eqnarray}

\section{Derivation of the Currents}\label{appendixB}

From the evolution equation $\frac{\partial}{{\partial}t}|\mathbb{P}_{\chi}\rangle=\mathbb{L}_{\chi}|\mathbb{P}_{\chi}\rangle$, we can define the characteristic function
\begin{equation}
\mathcal{Z}(\chi,t)=\langle\mathrm{1}|\mathbb{P}_{\chi}(t)\rangle=\langle\mathrm{1}|e^{\mathbb{L}_{\chi}t}|\mathbb{P}_{\chi}(0)\rangle,
\end{equation}
where $\langle\mathrm{1}|=(1,1,1,0,0)$ considering $\rho_{GG}+\rho_{LL}+\rho_{RR}=1$. In the long time limit, the cumulant generating function can be then expressed as~\cite{rj1, rj2}
\begin{equation}
\mathcal{G}(\chi)=\lim_{t\rightarrow\infty}\frac{1}{t}\mathcal{Z}(\chi,t)=\lambda_0(\chi),
\end{equation}
where $\lambda_0(\chi)$ is the eigenvalue of the operator $\mathbb{L}_{\chi}$, which has the largest real part and thus dominates the dynamics in the steady state.
The current is just the first order cumulant that is then obtained by the first order derivative
\begin{equation}
\mathcal{I}:=\frac{{\partial}\mathcal{G}(\chi)}{{\partial}(i\chi)}|_{\chi=0}=\frac{\partial\lambda_0(\chi)}{\partial(i\chi)}|_{\chi=0}=\left\langle\mathrm{1}\left|\left.\frac{\partial\mathbb{L}_{\chi}}{\partial(i\chi)}\right|_{\chi=0}\right|\mathbb{P}^{ss}\right\rangle.
\end{equation}
For the specific current calculation, $\chi=\chi_e$ gives the electron current,
and $\chi=\chi_p$ gives the photon flow.

Therefore, the electron current
is obtained as
\begin{eqnarray}
I_e/e&=&\left\langle\mathrm{1}\left|\left.\frac{{\partial}\mathbb{L}^{e}_{\chi_e}}{{\partial}(i\chi_e)}\right|_{\chi_e=0}\right|\mathbb{P}^{ss}\right\rangle  \\
&=&\Gamma_L\rho^{ss}_{LL}-\Gamma_{GL}\rho^{ss}_{GG}+2\Theta_{GL}\textrm{Re}[\rho^{ss}_{LR}],\nonumber
\end{eqnarray}
where $|\mathbb{P}^{ss}{\rangle}$ is the vector of the density matrix in steady state.
Similarly, the photon flow out of the environment can also be obtained as
\begin{eqnarray}~\label{Ip1}
I_p&=&-\left\langle\mathrm{1}\left|\left.\frac{{\partial}\mathbb{L}^{p}_{\chi_p}}{{\partial}(i\chi_p)}\right|_{\chi_p=0}\right|\mathbb{P}^{ss}\right\rangle\\
&=&-\frac{\gamma_p}{2}(\sin^2\theta{(\rho^{ss}_{LL}+\rho^{ss}_{RR})}\nonumber\\
&&+2\sin\theta(1+2n(\Lambda))\textrm{Re}[\rho^{ss}_{LR}]).\nonumber
\end{eqnarray}
Since the counting field counts the photon current into the reservoir, there is a minus sign for calculating the photon current out of the reservoir.

\end{appendix}

\begin{acknowledgements}
This work was supported by the National Science Foundation (NSF) (grant no. CHE-1112825) and Defense Advanced Research Projects Agency (DARPA) (grant no. N99001-10-1-4063).
C. Wang has been supported by Singapore-MIT Alliance for Research and Technology (SMART).
J. Ren acknowledges the auspices of the National Nuclear Security Administration of the U.S. DOE at LANL under Contract No. DE-AC52-06NA25396, through the LDRD Program.
J. Cao has been supported by the Center for Excitonics,
an Energy frontier Research Center funded by the U.S. Department of Energy,
Office of Science, Office of Basic Energy Science.  
\end{acknowledgements}

\end{document}